# Dependence of radio wave anomalous attenuation in the ionosphere on properties of spatial spectrum of irregularities

N.A. Zabotin, G.A. Zhbankov and E.S. Kovalenko

*Rostov State University, Rostov-on-Don, Russia*

**Abstract.** The paper presents theoretical study of the influence of shape of the spatial spectrum of small-scale ionospheric irregularities on anomalous attenuation of HF radio waves caused by multiple scattering. Several spectrum models are considered. Relative contribution of different wave number regions of the spectrum is investigated. The analysis of numerical calculations shows that the shape of the irregularity spectrum can be determined from the distribution of the signal attenuation on the Earth's surface.

**Introduction**

Anomalous attenuation of radio waves reflected from ionosphere which cannot be explained by collision loses is observed in vertical sounding of ionosphere both in natural conditions and in the heating experiments. Two mechanisms are known that may be responsible for anomalous attenuation. The first mechanism is known also as anomalous absorption. The reason of anomalous absorption is transformation of ordinary wave into plasma wave due to scattering from small-scale (characteristic scale of order of wavelength or less) irregularities. This mechanism plays important role in generation of artificial small-scale irregularities and attenuation of powerful pump wave of ordinary polarization and ordinary probe waves. It does not affect extraordinary wave. In natural conditions anomalous absorption is not important because the small-scale irregularities are usually suppressed.

The second mechanism is scattering from irregularities with characteristic scales in the range 100 m – 10 km. The scattering influences on the waves of both polarizations. In the conditions of non-disturbed mid-latitude ionosphere in may reach a value of 10 dB. It may be also important in heating experiments causing additional attenuation of ordinary wave and attenuation of extraordinary wave.

Theoretic description of anomalous attenuation of radio waves caused by scattering may be performed by the means of the theory of radiative transfer in random media [Bronin and Zabotin, 1992; Zabotin, et al., 1998a]. This approach was used to calculate angular and spatial distribution of radiation reflected from plane-stratified slab for the simple model of power-law spectrum of irregularities [Zabotin, et al., 1998a]. However, it is known that both natural and artificial irregularities are characterized by rather complex spectra with different behavior in different ranges of characteristic scales [Frolov et al, 1996; Szuszczewicz, 1987]. The behavior of spectra is not well known for the largest and the smallest scales. To determine parameters of real ionospheric irregularity spectra from the observations of anomalous attenuation it is important to know how the peculiarities of spectrum influence on anomalous attenuation of sounding signal.

## 1. Calculation of anomalous attenuation

Theoretic description of scattering from random irregularities in ionosphere is a rather difficult problem, because one has to take into account hyrotropy of the medium, regular refraction and the multiple character of scattering. If only the intensity of the wave is of interest, the scattering may be described by means of the radiative transfer equation [Bronin and Zabotin, 1992; Zabotin, et al., 1998a].

The ray path in plane-stratified medium may be unambiguously defined by invariant ray variables – polar and azimuth angles of arrival $\theta, \varphi$ at the point $\vec{\rho}$ on the certain base plane, parallel to the layer. It is natural to choose the surface of the Earth as the base plane for the case ionospheric propagation. The spatial and angular distribution $P_0(\theta, \varphi, \vec{\rho})$ of radiation energy reflected from the ionosphere is described by the radiation balance equation (RBE) [Zabotin, et al., 1998a]:

$$\frac{d}{dz} P(z, \vec{\rho}, \theta, \varphi) = \int Q(z; \theta, \varphi; \theta', \varphi') \cdot$$
$$\cdot \left\{ P(z, \vec{\rho} - \vec{\Phi}(z; \theta', \varphi'; \theta, \varphi), \theta', \varphi') - P(z, \vec{\rho}, \theta, \varphi) \right\} d\theta' d\varphi' \quad , \tag{1}$$

where $z$ is vertical coordinate,

$Q(z; \theta, \varphi; \theta', \varphi') = \sigma(\theta, \varphi; \theta', \varphi') C^{-1}(z, \theta, \varphi) \sin \theta' \left| \frac{d\Omega'_k}{d\Omega'} \right|$, $C(z; \theta, \varphi)$ is cosine of angle of inclination of ray path with invariant angles $\theta$ and $\varphi$; $\left| \frac{\partial(\alpha, \beta)}{\partial(\theta', \varphi')} \right|$ is Jacobean of transition from polar and azimuth angles of wave vector to invariant angles, $\sigma(z, \theta, \varphi, \theta', \varphi')$ is scattering cross-section. Vector function $\vec{\Phi}(z; \theta, \varphi; \theta', \varphi')$ describes the displacement of the point of arrival of the scattered ray with angle coordinates $\theta', \varphi'$, in relation to the point of arrival of incident ray with angle coordinates $\theta, \varphi$.

In the approximation of small-angle scattering in invariant angle variables (which is not identical to common small-angle scattering approximation) the solution of REB equation is

$$P(z, \vec{\rho}, \theta, \varphi) = P_0 \left[ \vec{\rho} - \vec{D}(z, 0; \theta, \varphi), \theta, \varphi \right], \tag{2}$$

where $P_0(\theta, \varphi, \vec{\rho})$ is spatial and angular distribution of radiation energy reflected from the ionosphere in absence of irregularities,

$$\vec{D}(z,0;\omega) = \int_0^z dz' \int d\theta' d\varphi' Q(z';\theta,\varphi,\theta',\varphi') \vec{\Phi}(z';\theta,\varphi,\theta',\varphi'). \tag{3}$$

According to (2) and (3) the scattering causes deformation of distribution of radiation energy, while the type of distribution function is not changed. If only the one ray with invariant angles $\theta_0(\vec{\rho})$, $\varphi_0(\vec{\rho})$ arrives in every point at the Earth surface $\vec{\rho}$ (as it takes place for the point source and frequencies below critical frequency) then the function $P_0$ has the following form

$$P_0(\vec{\rho},\theta,\varphi) = \tilde{P}_0(\vec{\rho})\delta[-\cos\theta + \cos\theta_0(\vec{\rho})]\delta[\varphi - \varphi_0(\vec{\rho})], \tag{4}$$

where the quantity $\tilde{P}_0(\vec{\rho})$ is proportional to energy flux through the point $\vec{\rho}$ in absence of scattering. Substitution of (4) into (2) gives:

$$\tilde{P}(\vec{\rho}) = \tilde{P}_0\left[\vec{\rho} + \vec{D}(\theta_1,\varphi_1)\right]\left|\frac{\partial(\rho_{0x},\rho_{0y})}{\partial(\theta,\varphi)}\right|\left|\frac{\partial(\rho_{0x} - D_x, \rho_{0y} - D_y)}{\partial(\theta,\varphi)}\right|^{-1}_{\substack{\theta=\theta_1,\\ \varphi=\varphi_1}}. \tag{5}$$

where $\vec{D}(\theta,\varphi) \equiv \vec{D}(z_0,0;\theta,\varphi)$. New angles of arrival $\theta_1$ and $\varphi_1$ may be found from the system of equations

$$\vec{\rho} = \vec{\rho}_0(\theta_1,\varphi_1) + \vec{D}(\theta_1,\varphi_1), \tag{6}$$

where $\vec{\rho}_0(\theta,\varphi)$ is the point of arrival at the base plane of the ray with invariant ray variables $\theta$ and $\varphi$ in absence of scattering.

According to (5), observer situated at point $\vec{\rho}$ will notice two effects caused by scattering: the change of angles of arrival and attenuation of intensity of received signal, which is determined as

$$L = 10\lg\frac{\tilde{P}(\vec{\rho})}{\tilde{P}_0(\vec{\rho})} \tag{7}$$

Formulae (5) – (7) are valid in general case when the effect of geomagnetic field on ray paths and scattering cross-section is taken into account. However, in [Zabotin et al., 1998b] it was found that taking into account the influence of geomagnetic field on ray paths and scattering cross-section does not lead to considerable changes in calculated anomalous attenuation. That is why further in this paper we will use isotropic plasma approximation for scattering cross-section and for determination of ray paths as it considerably simplifies numerical calculations. The effect of geomagnetic field in this case is taken into account only through the spectrum of irregularities which is considered to be anisotropic (see the next Section). Detailed description of numeric calculation of the anomalous attenuation in this approximation may be found in [Zabotin, et al., 1998a].

## 2. Anomalous attenuation for simple models of ionospheric irregularities spectrum

The simplest model of the spectrum of the ionospheric irregularities uses the fact that ionospheric irregularities at least in the band of scales responsible for scattering are strongly stretched along geomagnetic field lines. Because the outer scale of the irregularities in the direction collinear with geomagnetic field lines is several orders greater than in orthogonal direction we may consider that the irregularities are infinitely stretched and characterized by the spectrum of the following form

$$\Phi(\vec{\kappa}) = C_A \left(1 + \kappa_\perp^2 / \kappa_{0\perp}^2\right)^{-\nu/2} \delta\left(\kappa_\parallel\right) \tag{8}$$

where $\kappa_\perp$ and $\kappa_\parallel$ are orthogonal and parallel to geomagnetic field lines components of wave vector of irregularities $\vec{\kappa}$, $\kappa_{0\perp} = 2\pi/L_m$, $L_m$ is upper scale of the spectrum, $\delta(x)$ is delta-function. The spectrum is normalized at the value of structural function

$$D_N\left(\vec{R}\right) = \left\langle \left[\delta_N\left(\vec{r} + \vec{R}\right) - \delta_N\left(\vec{r}\right)\right]^2 \right\rangle \equiv \delta_R^2,$$

where $\delta_N(\vec{r}) = \Delta N/N$ for the orthogonal scale $R = 1$ km. $C_A$ is a normalization constant which depends on the spectrum parameters:

$$C_A = \delta_R^2 \frac{\Gamma(\nu/2)}{2\pi\kappa_{0\perp}^2} \left[\Gamma\left(\frac{\nu-2}{2}\right) - 2\left(\frac{R\kappa_{0\perp}}{2}\right)^{\frac{\nu-2}{2}} K_{\frac{\nu-2}{2}}(R\kappa_{0\perp})\right]^{-1},$$

$\Gamma(x)$ is the Gamma-function, and $K_\beta(z)$ is the Macdonald function [Abramowitz and Stegun, 1972].

The example of spatial distribution of anomalous attenuation of the reflected sounding signal at the Earth's surface is shown at Fig. 1 [Zabotin, et al., 1998a]. Calculation were done for the latitude of N. Novgorod (the inclination of geomagnetic field $\gamma = 19°$) for the following set of parameters: $L_m = 10$ km, $\nu = 2.5$, $\delta_R = 0.003$, sounding frequency $f = 4$ MHz. The altitude dependence of electron density was modeled by linear profile with the beginning at the altitude of $h_0 = 150$ km and height of reflection of 4 MHz signal at 250 km (thickness of the slab $H = 100$ km). This model of electron density profile was used also in other calculations described below. The same distribution of anomalous at the Earth's surface in the vicinity of sounding station but for $\nu = 3.9$ is presented at Fig. 2. In both cases the distribution of anomalous attenuation is symmetrical in relation to the plane of magnetic meridian. The whole distribution may be divided into two regions. The first region has elliptic form and contains the origin of the coordinate system, where the sounding station is settled. In this region the strong ( ~10 dB) attenuation is observed. In the second outer region the attenuation is near zero or even negative. This is not surprising because when the collisional absorption is not taken into account the total energy of the signal is not changed but only redistributed in the space due to scattering.

The other possible model of spectrum of irregularities is the model of infinitely stretched irregularities with Gauss-law spectrum in orthogonal direction. According to the modern concept of ionospheric irregularities such spectra are not realistic. However it may be useful to understand how the shape of the spectrum influences the distribution of anomalous attenuation in the vicinity of sounding station. The spectrum is defined by the following expression:

$$\Phi(\vec{\kappa}) = C_G \exp\left[-\frac{(\kappa_\perp - \kappa_{0\perp})^2}{\Delta\kappa_\perp^2}\right]\delta(\kappa_\|), \qquad (9)$$

where $\kappa_{0\perp}$ is the wave number corresponding to the maximum of the spectrum, $\Delta\kappa_\perp$ is the half-width of the spectrum: $\Phi(\kappa_\perp = \kappa_{0\perp}) = e \cdot \Phi(\kappa_\perp = \kappa_{0\perp} + \Delta\kappa_\perp)$, where $e$ is the base of natural logarithms, $C_G$ is a normalization constant:

$$C_G = \frac{\delta_R^2}{4\pi \cdot I_{str}}, \text{ where } I_{str} = \int_0^\infty \kappa_\perp \exp\left[-\frac{(\kappa_\perp - \kappa_{0\perp})^2}{\Delta\kappa_\perp^2}\right] \left[1 - J_0(\kappa_\perp R)\right] d\kappa_\perp,$$

and dispersion of irregularities $\sigma_N^2$ is connected with this constant by the formula

$$C_G = \frac{\sigma_N^2}{4\pi \cdot I_{disp}}, \text{ where } I_{disp} = \int_0^\infty \kappa_\perp \exp\left[-\frac{(\kappa_\perp - \kappa_{0\perp})^2}{\Delta\kappa_\perp^2}\right] d\kappa_\perp.$$

When $\kappa_{0\perp} \gg \Delta\kappa_\perp$ one has $I_{disp} \approx \sqrt{\pi}\kappa_{0\perp}\Delta\kappa_\perp$.

Distribution of anomalous attenuation for this Gaussian model of spectrum for $R = 1$ km, $\delta_R = 0.002$, sounding frequency $f = 4$ МГц, $\gamma = 19°$, $L_{0\perp} = 2\pi/\kappa_{0\perp} = 2$ km, $\Delta\kappa_\perp = 0.3 \cdot \kappa_{0\perp}$ is shown at Fig. 3. The important peculiarity of this distribution is that it has two maxima of attenuation symmetrical to the plane of magnetic meridian and these maxima are not in the origin of coordinate system. It may be explained by the suppression of large scales in Gaussian spectrum. This point of view is supported by the results of calculations for the model of power-law spectrum with the cut-off of large scales which will be described in the next Section.

## 3. Combined models of spectrum

The simple models of spectrum of ionospheric irregularities are not quite adequate to the real spectra of ionospheric irregularities. In the case of artificial irregularities generated by the heating of ionosphere by the powerful HF wave the spectrum as a rule has a hump in the band of scales 0.5 - 1 km. Spectrum of such type may be modeled as a combination of two simple power-law type spectra:

$$\Phi(\vec{\kappa}) = \left\{ C_{A_1}\left(1 + \kappa_\perp^2/\kappa_{0\perp_1}^2\right)^{-\nu_1/2} + C_{A_2}\left(1 + \kappa_\perp^2/\kappa_{0\perp_2}^2\right)^{-\nu_2/2} \right\} \delta(\kappa_\parallel), \qquad (10)$$

where the normalization constants are of the same form as in (8):

$$C_{A_i} = \delta_{R_i}^2 \frac{\Gamma(\nu_i/2)}{2\pi\kappa_{0\perp_i}^2}\left[\Gamma\left(\frac{\nu_i-2}{2}\right) - 2\left(\frac{R_i\kappa_{0\perp_i}}{2}\right)^{\frac{\nu_i-2}{2}} K_{\frac{\nu_i-2}{2}}(R_i\kappa_{0\perp_i})\right]^{-1}, \quad i = 1,2.$$

At Fig. 5a the construction of the orthogonal component of the spectrum for $\nu_1 = 2.5, \delta_{R_1} = 0.003$ and $\nu_2 = 3.9, \delta_{R_2} = 0.005$ is demonstrated. At Fig. 5b the frequency dependence of anomalous attenuation at vertical sounding is shown for each of the spectra and for total spectrum. It must be noted that anomalous attenuation is not additive function in relation to parts of combined spectrum. Anomalous attenuation at vertical sounding as a function of the ratios of structural functions $\delta_{R_2}/\delta_{R_1}$ and outer scales $L_2/L_1 = \kappa_{0\perp_1}/\kappa_{0\perp_2}$ of partial spectra is plotted at Fig. 6.

Results of calculation of anomalous attenuation for such spectrum and the same set of parameters as for Fig. 1 are shown at Fig.7. It is easy to note that the resulting distribution of anomalous attenuation is a composition of distributions for each of partial spectra. In this case total distribution on the surface of Earth may be divided into three regions: the region of high attenuation (corresponding to mutual effect of partial spectra), the region of medium attenuation and the region of slight attenuation. So, the stepped form of distribution is the indication of twist in the spectrum and this fact may be useful for diagnostics of the spectrum.

All the models of power-law spectrum discussed above are based on supposition that in the region of large scales the spectrum is saturated. In other words, for the waves numbers $\kappa_\perp > \kappa_{0\perp}$ the spectrum is described by power law and for small wave numbers $\kappa_\perp \ll \kappa_{0\perp}$ the spectrum is constant. This may not take place for real spectra because the upper scale of irregularities is not infinite. The upper scale of irregularities is limited at least by the sizes of the medium and it is natural to suppose that irregularities with the scales greater than outer scale are destroyed by some dissipation processes. It means that the realistic spectrum model must include some cut-off at the large scales. Such cut-off may be introduced with the help of Gauss function:

$$\Phi(\vec{\kappa}) = C_A\left(1 + \kappa_\perp^2/\kappa_{0\perp}^2\right)^{-\nu/2} \exp\left(-K_{obr} \cdot \kappa_{0\perp}^2/\kappa_\perp^2\right) \delta(\kappa_\parallel). \qquad (11)$$

At Fig. 8 the orthogonal component of such spectrum as a function of $\kappa_\perp$ for different $K_{obr}$ and dependence of anomalous attenuation on $K_{obr}$ are plotted. The distributions of anomalous attenuation of reflected signal in some vicinity of sounding station for the following set of parameters $K_{obr} = 2$, $f = 4$ MHz, $\nu = 2.5$, $\delta_R = 0.3\%$, $H = 100$ km, $h_0 = 150$ km, $L_m = 10$ km, $R = 1$ km and inclination angles (different latitudes) are plotted at Fig. 9. It is easy to note that distributions in this case are similar to the one for gauss spectrum model. In both cases we have two symmetrical maximums instead one in the origin of coordinate system. This fact proves our supposition that such behavior is a result of cut-off of large scales.

## 4. Contribution into anomalous attenuation from different wave numbers in the spectrum

It is interest to determine the contribution from different regions of wave numbers in the spectrum. However, it is difficult to do it directly because of nonlinear relation of the spectrum and anomalous attenuation as it is clear from formulae (5) – (7). However it may be performed relatively simply using the following indirect way. Let us introduce spectrum combined from power-law and Gauss spectra:

$$\Phi(\vec{\kappa}) = \left\{ C_A \left(1 + \frac{\kappa_\perp^2}{\kappa_{0\perp}^2}\right)^{-\nu/2} + C_G \left[ -\frac{(\kappa_\perp - \kappa_{G0\perp})^2}{\Delta \kappa_\perp^2} \right] \right\} \delta(\kappa_\parallel). \qquad (12)$$

The shape of this spectrum is illustrated by the Fig. 10. Varying the position of the maximum of Gauss spectrum $\kappa_{G0\perp}$, one may obtain required dependence. In our case it is more interesting to study the dependence of derivatives $\frac{\partial L}{\partial(\Delta\kappa_\perp)}$ and $\frac{\partial L}{\partial \sigma_N^2}$ on $\kappa_{G0\perp}$, what gives more exact information about the contribution of different regions of spectrum into anomalous attenuation. The results of calculations are presented at Fig. 11 – 13. The values of these derivatives show the value of the contribution to the effect from irregularities with a given $\kappa_{G0\perp}$. Let us consider that the region of $\kappa_{G0\perp}$, corresponding to half-width of the curve $\frac{\partial L(\kappa_{G0\perp})}{\partial \sigma_N^2}$ at Fig. 11b gives main contribution to the effect. Then from Fig. 11b one may conclude that on the latitude of Sura heating facility ($\gamma = 19^o$) the effect of anomalous attenuation mainly arises from irregularities with scales 100-600 m.

## Conclusion

Distributions of anomalous attenuation of radio wave reflected from ionosphere in the vicinity of sounding station, calculated in the paper, demonstrate specific dependencies on different parameters of spatial spectrum of irregularities. This fact allows one to use the measurements of anomalous attenuation in a sufficient number of points at the Earth's surface in the vicinity of sounding station for qualitative determination of spectrum shape and parameters. In particular, from the distribution of anomalous attenuation it is possible to determine if the spectrum has a bend at a certain wave number and to estimate this wave number, or to determine the behavior of the spectrum at large scales. This information helps to choose an adequate model of spectrum of random ionospheric irregularities and to determine the parameters of spectrum like spectral index, outer scale and structural function. i.e. to solve the inverse problem.

*Acknowledgments.* The work was supported by the Russian Foundation of Basic Research under grant No. 99-02-17525.

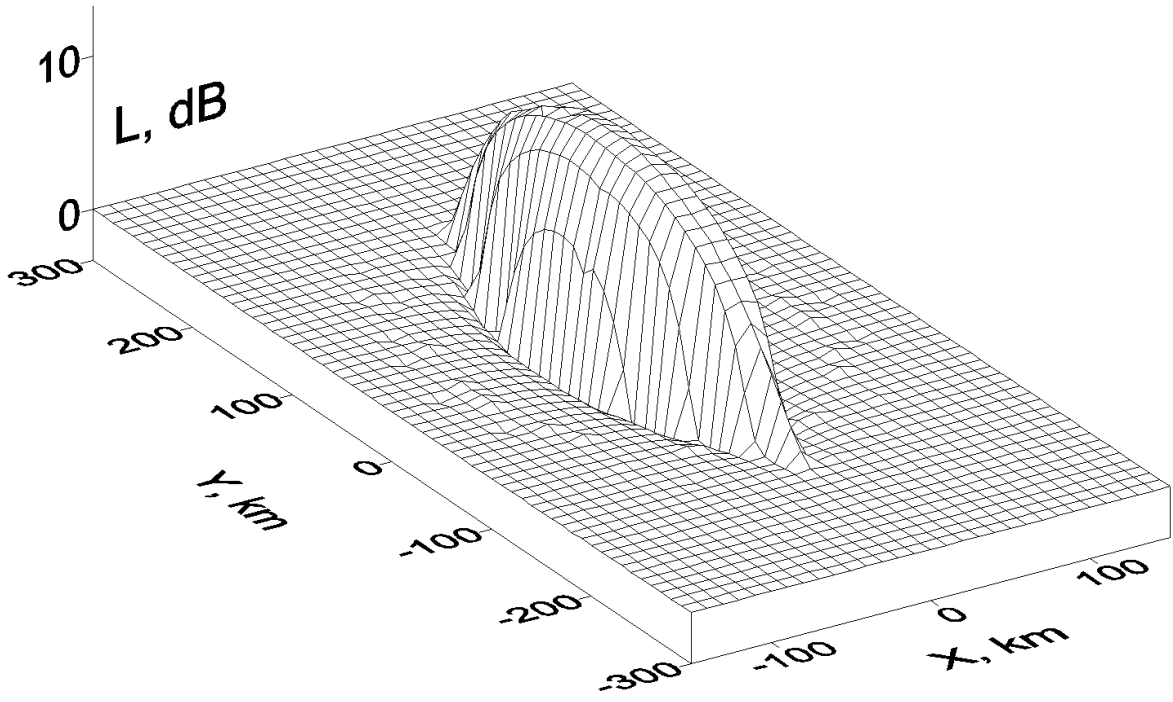

Fig. 1.

Distribution of anomalous attenuation in the vicinity of sounding station for simple power-law spectrum. Spectral index $\nu = 2.5$.

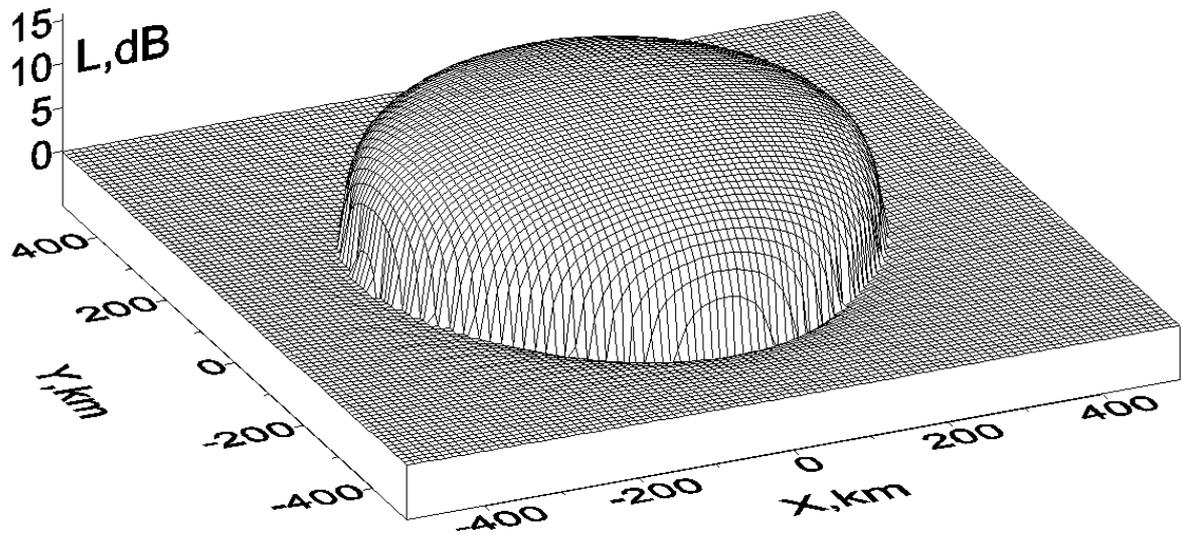

Fig.2.

Distribution of anomalous attenuation in the vicinity of sounding station for simple power-law spectrum. Spectral index $\nu = 3.9$.

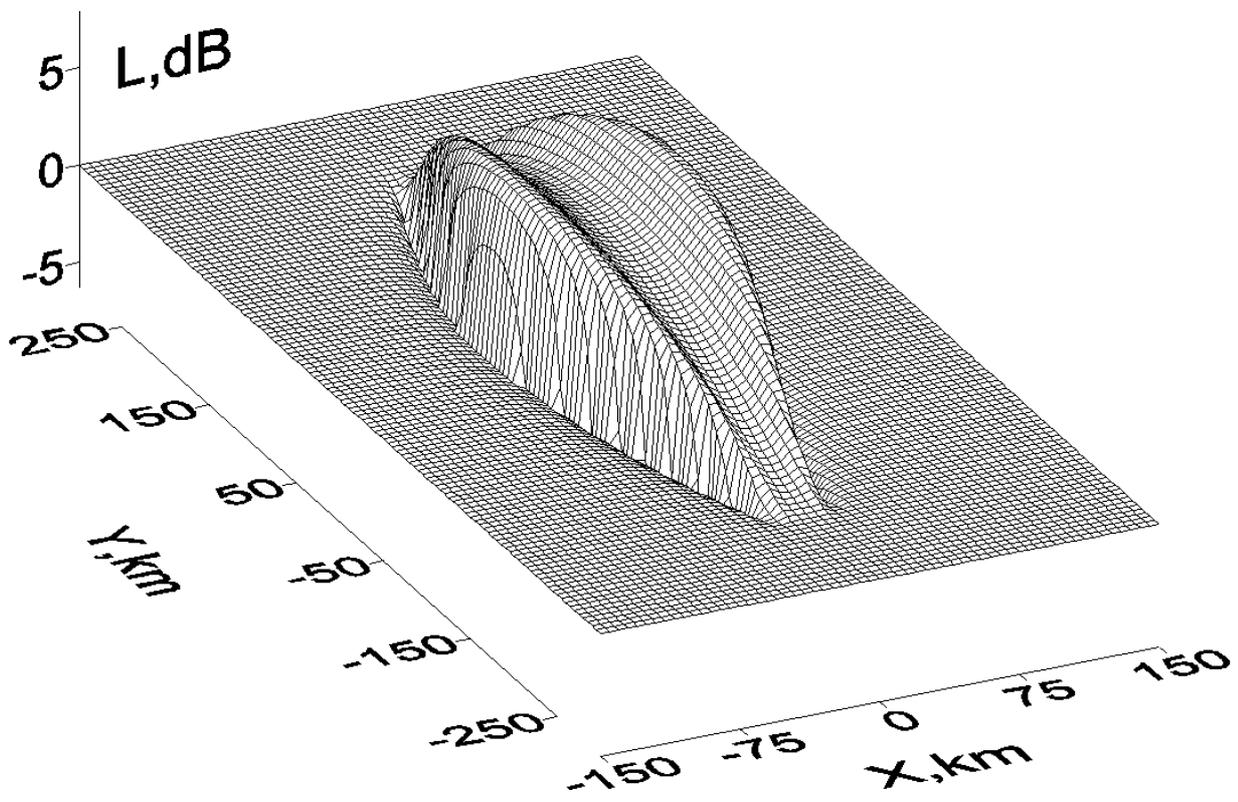

Fig. 3.

Spatial distribution of anomalous attenuation in the vicinity of sounding station for Gaussian spectrum.

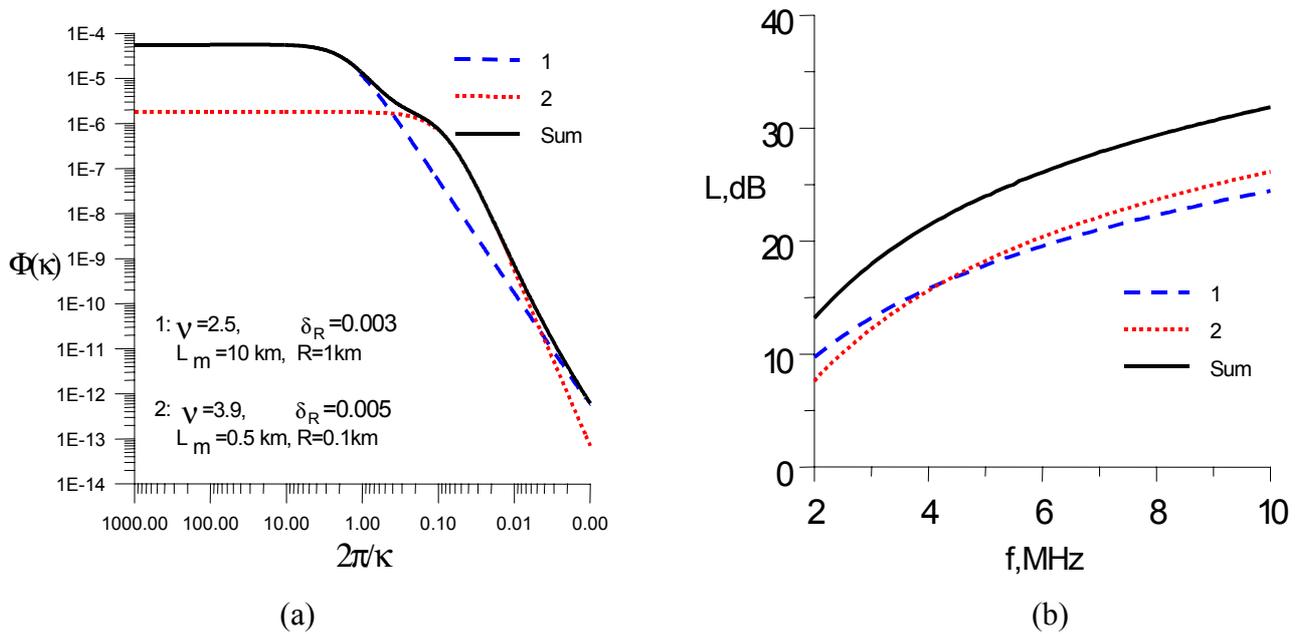

(a)                                                                                 (b)

Fig. 5.

Combimed spectrum.

a) Dependence of orthogonal component of the spectrum on wave number of irregularities.

b) Anomalous attenuation at vertical sounding as a function of sounding frequency for combined spectrum.

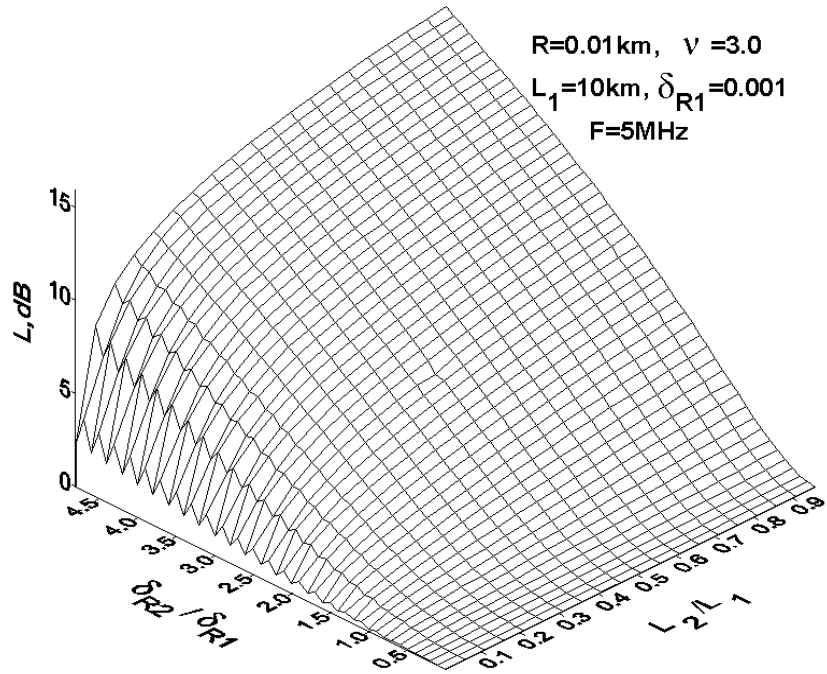

Fig. 6.

Dependence of anomalous attenuation for combined spectrum on the ratios of structural functions $\delta_{R_2}/\delta_{R_1}$ and outer scales $L_2/L_1 = \kappa_{0\perp_1}/\kappa_{0\perp_2}$ of partial spectra.

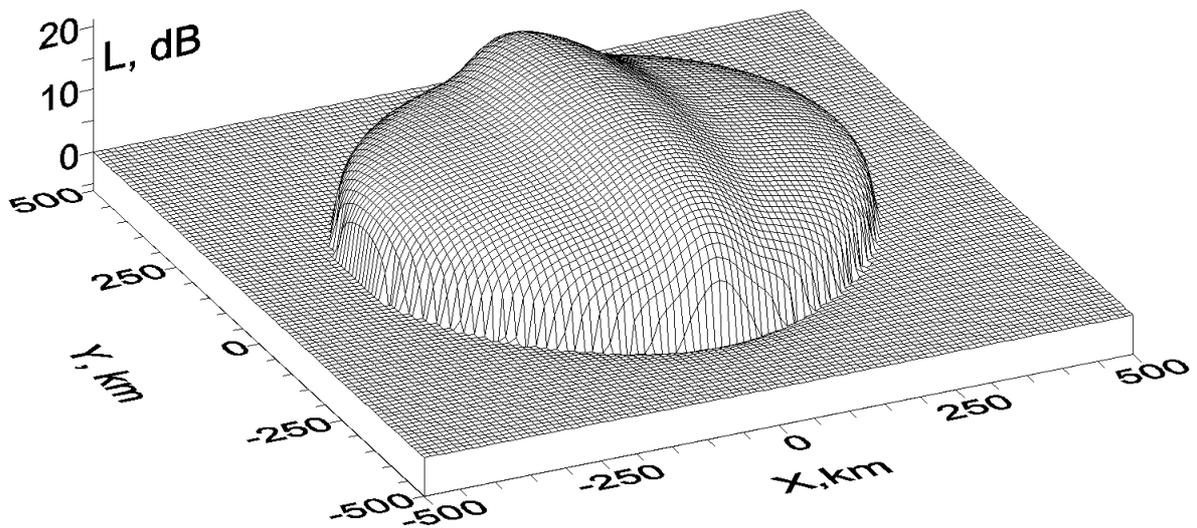

Fig. 7.

Spatial distribution of anomalous attenuation in the vicinity of sounding station for combined spectrum.

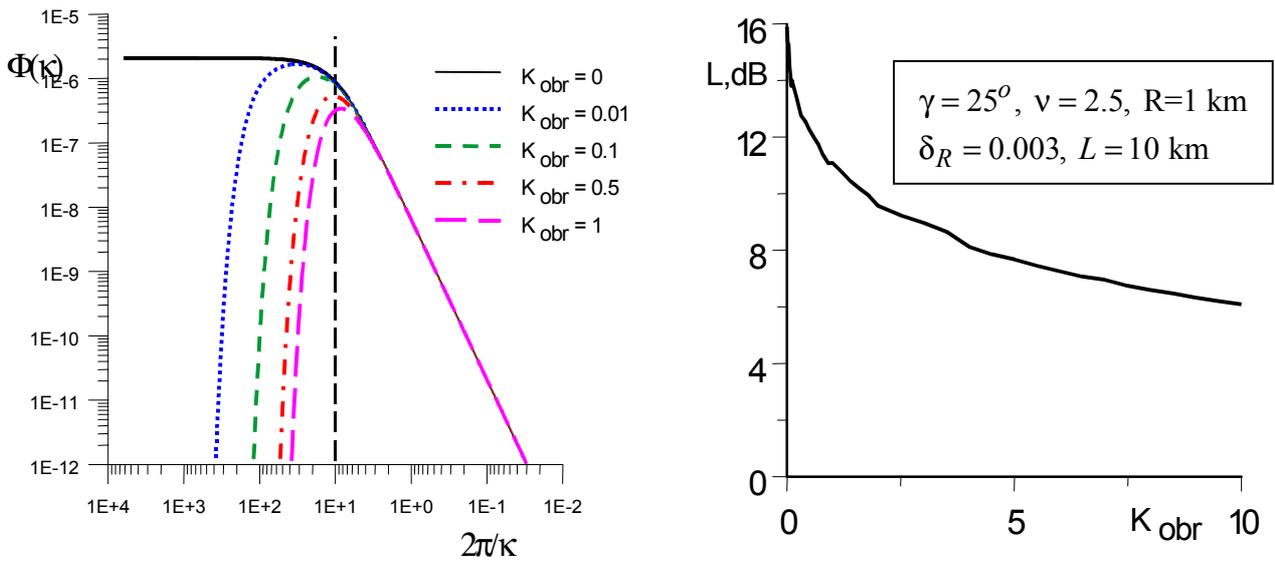

Fig.8.

Spectrum with cut-off at the large scales.

a) Dependence of orthogonal component of the spectrum on wave number of irregularities for different cut-off wavenumbers $K_{obr}$.

b) Anomalous attenuation at vertical sounding as a function of cut-off wave number $K_{obr}$.

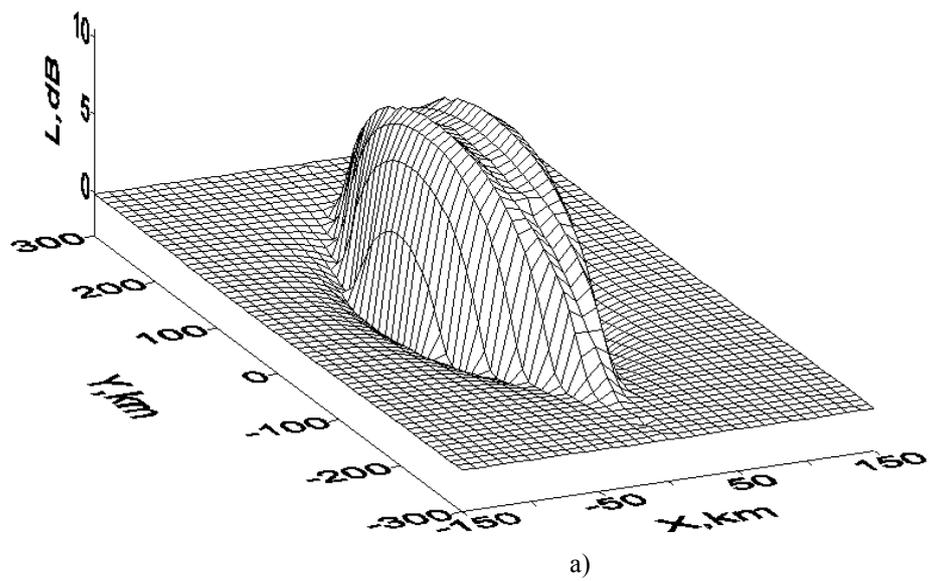

a)

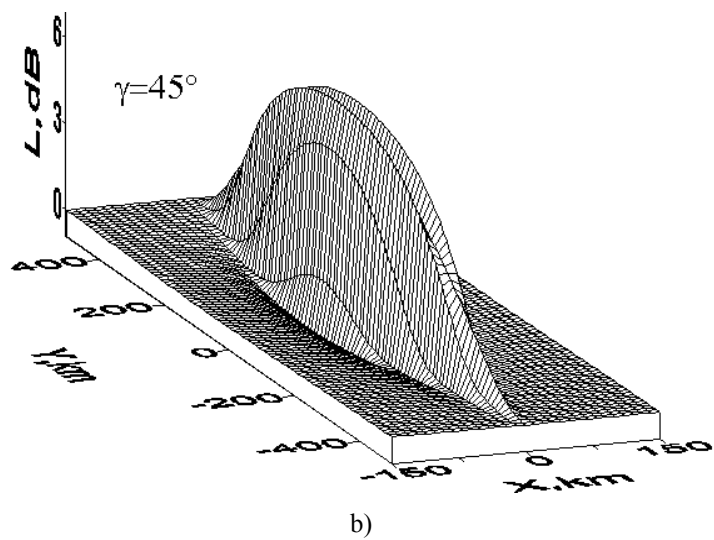

b)

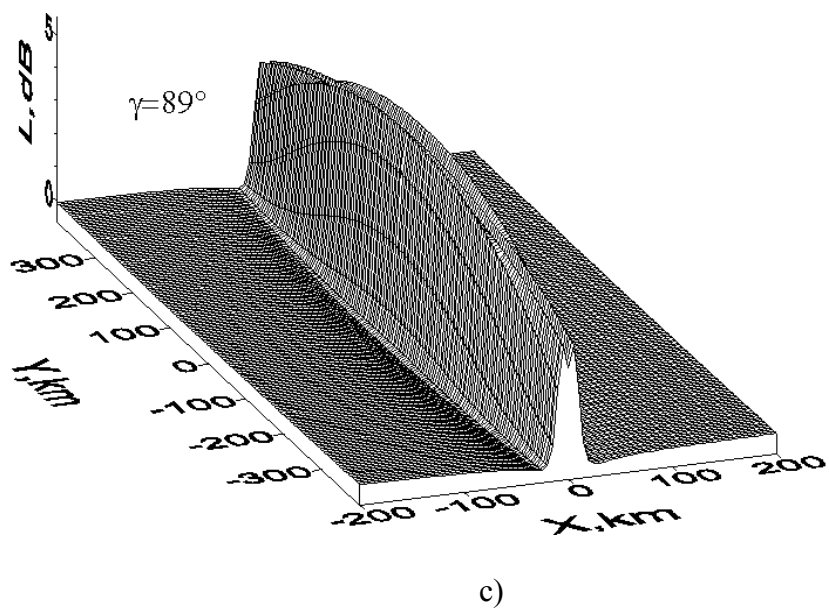

c)

Fig.9.

Spatial distribution of anomalous attenuation in the vicinity of sounding station for the spectrum.with cut-off.

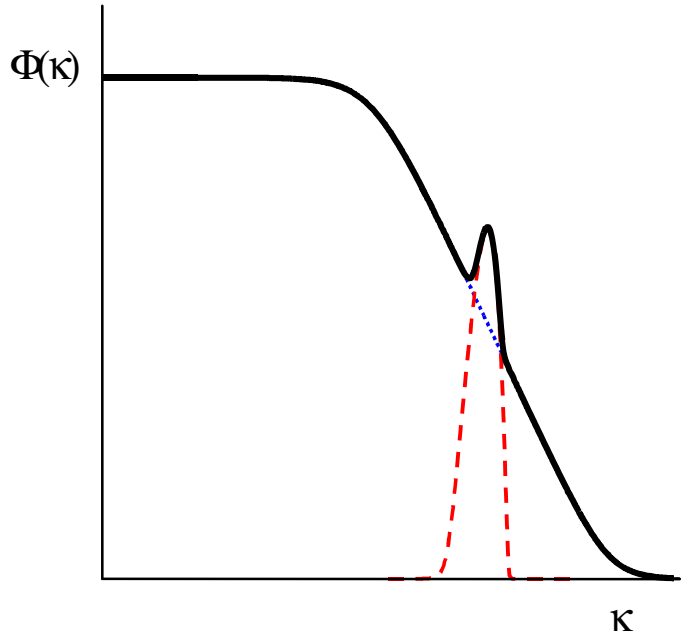

Fig.10.

Illustration of the method for determination of contribution to anomalous attenuation from different wave numbers in the spectrum .

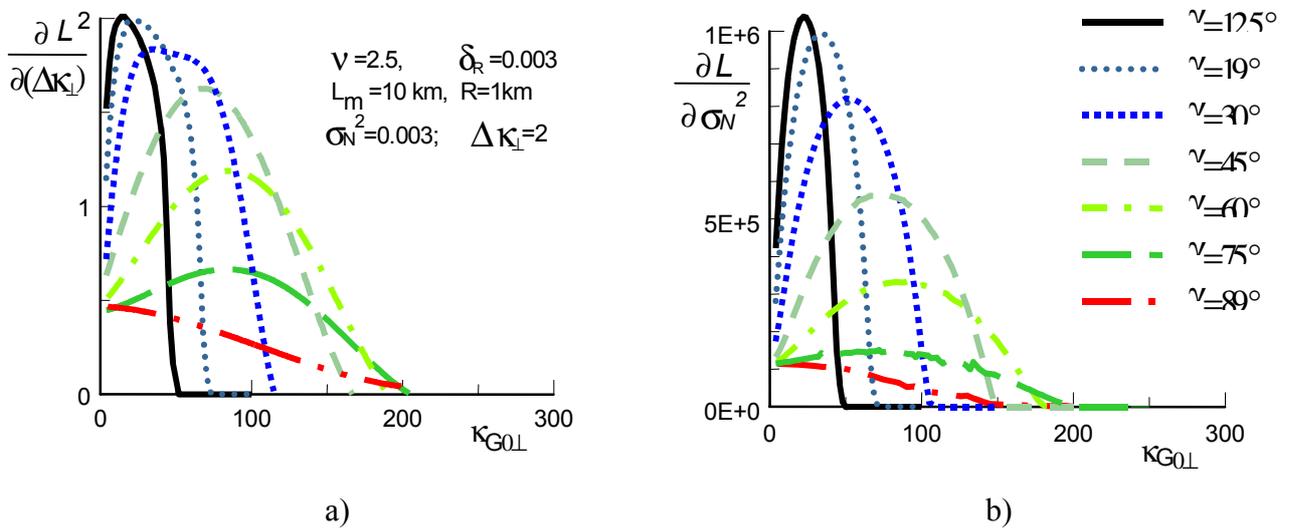

a)                                                                   b)

Fig. 11.

Dependencies of quantities $\dfrac{\partial L}{\partial (\Delta \kappa_\perp)}$ (a ) and $\dfrac{\partial L}{\partial \sigma_N^2}$ (b) on $\kappa_{G0\perp}$ for various latitudes.

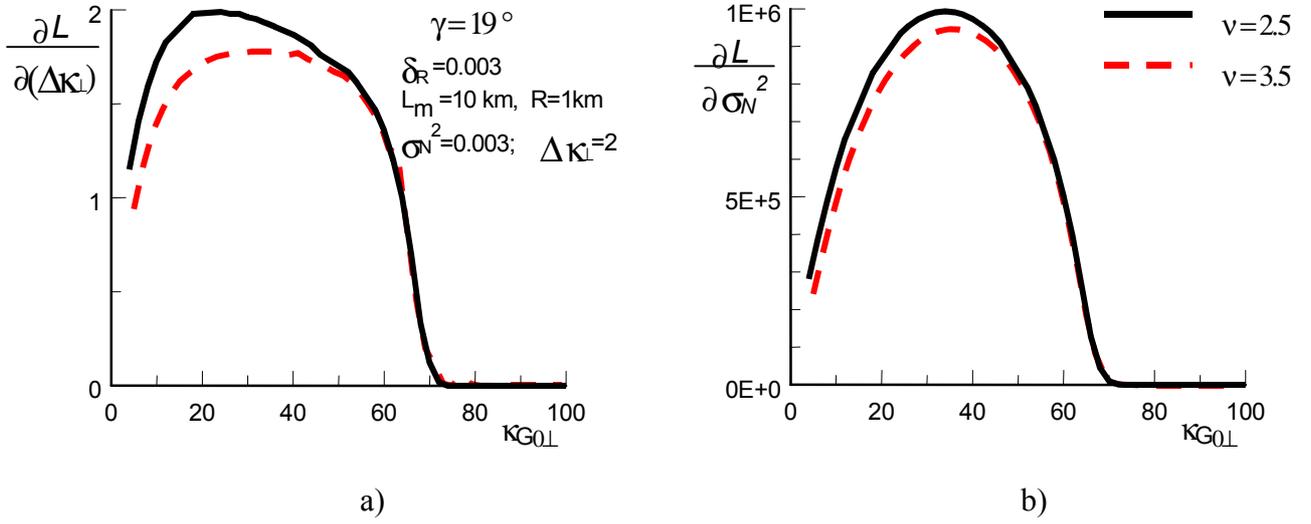

Fig. 12.

Dependencies of quantities $\dfrac{\partial L}{\partial(\Delta\kappa_\perp)}$ (a) and $\dfrac{\partial L}{\partial\sigma_N^2}$ (b) on $\kappa_{G0\perp}$ for high latitudes.

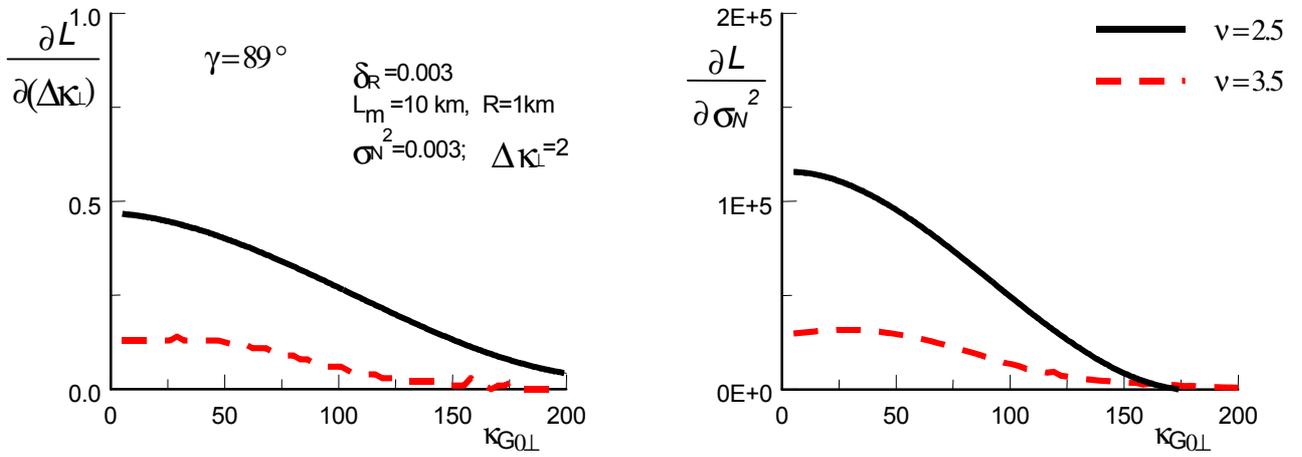

Fig. 13.

Dependencies of quantities $\dfrac{\partial L}{\partial(\Delta\kappa_\perp)}$ (a) and $\dfrac{\partial L}{\partial\sigma_N^2}$ (b) on $\kappa_{G0\perp}$ for low latitudes.